\documentclass[a4paper,11pt]{article}
\usepackage{pos}
\usepackage{graphicx,amsmath,color}
\usepackage{relsize}
\usepackage{enumerate}
\usepackage{slashed}
\usepackage{multirow}
\usepackage[normalem]{ulem}

\def\beq{\begin{equation}}
\def\eeq{\end{equation}}
\def\bea{\begin{eqnarray}}
\def\eea{\end{eqnarray}}

\def\bit{\begin{itemize}}
\def\eit{\end{itemize}}

\def\l{\left}

\def\baa{\begin{array}}
\def\eaa{\end{array}}

\def\simgt{\mathrel{\lower2.5pt\vbox{\lineskip=0pt\baselineskip=0pt
           \hbox{$>$}\hbox{$\sim$}}}}
\def\simlt{\mathrel{\lower2.5pt\vbox{\lineskip=0pt\baselineskip=0pt
           \hbox{$<$}\hbox{$\sim$}}}}

\def\bfc{\begin{figure}\begin{center}}
\def\efc{\end{center}\end{figure}}

\definecolor{chromeyellow}{rgb}{1.0, 0.65, 0.0}
\definecolor{darkcoral}{rgb}{0.8, 0.36, 0.27}
\definecolor{cadmiumgreen}{rgb}{0.0, 0.42, 0.24}

\title{DW-genesis: generating the baryon number from domain walls}

\author*[a]{Miguel Vanvlasselaer}

\affiliation[a]{Theoretische Natuurkunde and IIHE/ELEM, Vrije Universiteit Brussel,
\& The International Solvay Institutes, Pleinlaan 2, B-1050 Brussels, Belgium }

\emailAdd{miguel.vanvlasselaer@vub.be}

\abstract{We show that axionic domain walls, when they couple to the lepton number, can generate a net baryon and lepton number through the mechanism of spontaneous baryogenesis. In this talk, we study systematically the baryon asymmetry produced by these domain walls (DWs) at annihilation, and refer to this process as \emph{DW-genesis}. We find that the baryon number is maximised when the DW network collapses  approximately
at the moment when the $L$-violating interaction decouples. We explore the expected gravitational wave signal from the DW network annihilation and the prospects for detecting it, but conclude that successful DW-genesis is typically in tension with observable GW signal. Morever, we briefly discuss a possible suppression induced by the cancellation between the 
asymmetry created by ``opposite'' 
axionic domain walls attached to the string.  }

\FullConference{9th Symposium on Prospects in the Physics of Discrete Symmetries (DISCRETE2024)\\
 2–6 Dec 2024\\
Ljubljana, Slovenia\\}

\begin{document}
\maketitle

\section{Introduction}

The numerical value of the baryon asymmetry of the universe (BAU), obtained via big bang nucleosynthesis\cite{Fields:2019pfx} and the latest CMB measurements\cite{Planck:2015fie}, is given by
\bea 
\label{eq:observed_BAU}
Y_{\Delta B} \equiv \frac{n_{B}-n_{\overline{B}}}{s} \bigg|_{0} \approx (8.69 \pm 0.22) \times 10^{-11} \, . 
\eea 
where $n_{B, \overline{B}}$ and $s$ are the number densities of baryons, antibaryons and entropy evaluated at present time, respectively. Within the inflationary paradigm, such imbalance cannot be explained by initial conditions and requires a further baryogenesis mechanism. Baryogenesis, or baryogenesis via leptogenesis are dynamical phenomena that are able to produce a non-vanishing abundance of baryons dynamically. On the other hand, if such a mechanism involved scales much higher than the EW scale, it becomes extremely challenging to probe with any terrestrial probe. 

While some realisations of electroweak baryogenesis\cite{Kuzmin:1985mm, Shaposhnikov:1986jp},  mesogenesis \cite{Elor:2018twp}, low-scale leptogenesis\cite{Granelli:2022eru} and GUT baryogenesis\cite{PhysRevD.18.4500}, can be probed by low-energy experiments, a lot of such baryogenesis realisations remain desperately out of experimental reach, including high scale leptogenesis\cite{FUKUGITA198645}. Recently, a broad theoretical effort was undergone to relate baryogenesis realisation to possible gravitational wave background\cite{Cline:2020jre, Lewicki:2021pgr, Azatov:2021irb, Chun:2023ezg}. In this paper, we present a new baryogenesis realisation which takes advantage of the presence of axionic domain walls in the early universe and naturally induces gravitational waves: DW-genesis\cite{Daido:2015gqa,Mariotti:2024eoh} and discuss if it can be probed.

\section{Domain wall formation and their GW signal}

Domain walls are topological defects that arise when a discrete symmetry is spontaneously broken in the plasma\cite{Kibble:1976sj,Vilenkin:2000jqa, Saikawa:2017hiv}. In the case of \emph{axion} models, the discrete symmetries emerge from anomalous global symmetries\cite{Peccei:1977hh, Peccei:1977ur, Weinberg:1977ma, Wilczek:1977pj} and the low energy theory can be described by the lagrangian
\bea 
\label{eq:Lagaxion}
\mathcal{L} =  \partial_\mu \Phi^\dagger \partial^\mu \Phi - \lambda \bigg(|\Phi|^2 - v_a^2/2 \bigg)^2 - V(a) \, .
\eea 
Here we denote $\Phi = \rho/\sqrt{2} \exp\left(ia/v_a\right)$ with $\rho$ the radial mode and $a$ the axion. The last term in Eq.\eqref{eq:Lagaxion} is the axion potential induced by the mixed anomaly between the global $U(1)$ and a confining dark gauge group (e.g. a dark $SU(N)$). This confining sector explicitly breaks the $U(1)$ symmetry to a $\mathbb{Z}_{N_{\rm DW}}$ discrete symmetry.  
For a temperature below the confining scale, $T \lesssim \Lambda \simeq \sqrt{m_a f_a}$, the axion potential takes the form  
\bea 
\label{eq:potential_axion}
V(a) = \Lambda^4 \bigg(1- \cos \left(\frac{ a N_\text{DW}}{v_a}\right)\bigg)= m_a^2 f_a^2\bigg(1- \cos \left(\frac{ a }{f_a}\right)\bigg) , \qquad m_a = \Lambda^2/f_a\,,
\eea 
with the axion defined in the range $a \in [0,2\pi v_a)$,
and with the axion decay constant given by $f_a \equiv v_a/N_\text{DW}$. After the confinement of the dark gauge group, the DWs, which are kink solutions interpolating between the different degenerate vacua of the  $\mathbb{Z}_{N_{\rm DW}}$ discrete symmetry, form and organise in a complex system of DWs, a DW network.
The network quickly approaches (within a few Hubble time) an attractor solution, called the scaling regime \cite{Hiramatsu:2012sc, Ryden:1989vj, Hindmarsh:1996xv, Garagounis:2002kt, Oliveira:2004he, Avelino:2005pe, Leite:2011sc}, where  
the average number of DWs per Hubble patch is a constant of order one and the average velocity of the individual DW is mildly relativistic. 

The energy of a single
kink solutions \cite{Vilenkin:2000jqa} is characterized by the energy density per unit surface, i.e. the tension, $
\sigma\simeq 8 m_a f_a^2 $. In the scaling regime, the energy density of the DW network scales like $ \rho_{\rm DW} \sim \sigma H$ and tends to dominate the energy budget of the early universe \cite{Zeldovich1974CosmologicalCO, Sikivie:1982qv}. This domination of the universe energy occurs at the temperature  when
$\rho_{\rm DW} \approx \rho_{\rm rad}$ and is given by $
T_{\rm dom} = \left( 
10/\pi^{2} g_* \right)^{1/4} \sqrt{\sigma/M_{\rm pl}} $, where $M_{\rm pl}$ is the reduced Planck mass. DW domination can be avoided if a small energy difference $\Delta V$ biases the vacua mapped by the discrete symmetry \cite{Sikivie:1982qv, Gelmini:1988sf} and induces the collapse of the DW network at a temperature
$
    T_\text{ann} = \left(90/\pi^2 g_*\right)^{1/4}\sqrt{\Delta V M_\text{pl}/\sigma}\,.
$
Requiring that the universe is always radiation dominated implies that $T_{\rm ann} \gtrsim T_{\rm dom} $.

During the scaling regime of the DW network, gravitational waves (GWs) are produced by the movement of the DWs \cite{Saikawa:2017hiv,Hiramatsu:2010yz,Hiramatsu:2012sc,Hiramatsu:2013qaa} 
(see also \cite{Kitajima:2023kzu,Kitajima:2023cek,Ferreira:2023jbu,Chang:2023rll,Li:2023gil} 
for recent works). 
This emission finishes when the network collapses at $T_{\rm ann}$. The gravitational wave signal is  parameterized by the energy fraction 
    \bea 
    \Omega_{\rm GW}(t, f) = \frac{1}{\rho_c(t)} \bigg(\frac{d \rho_{\rm GW}(t)}{d \ln f}\bigg) \, ,
    \eea 
where $\rho_c(t) = 3 M_{\rm pl}^2 H^2$ is the critical density of the universe at time $t$. The GW spectrum can be described by a power broken law with 
\bea 
\Omega_{\rm GW}(t, f) = \Omega^{\rm peak}_{\rm GW} (t) \times 
\begin{cases}
(f/f_{\rm peak})^{3} \qquad f < f_{\rm peak}
\\
(f/f_{\rm peak})^{-1}  \qquad f > f_{\rm peak}
\end{cases} \, , \qquad \Omega_{\rm GW}^\text{peak}(t) = \frac{\tilde\epsilon_{\rm GW} G \mathcal{A}^2 \sigma^2}{\rho_c(t)}
\eea 
where $f_{\rm peak}$ is the frequency at which the GW signal is maximal. Here $\mathcal{A} = 0.7$ and  $\tilde\epsilon_{\rm GW} = 0.7$ taken from numerical simulations performed in \cite{Hiramatsu:2013qaa}. 
After redshifing to today, the amplitude of the GW signal reads
\bea 
\label{eq:GWfromDW}
\Omega_{\rm GW}(T)\approx 2.34 \times 10^{-6} \tilde \epsilon_{\rm GW} \mathcal{A}^2\bigg(\frac{g_\star (T)}{10}\bigg) \bigg(\frac{g_{s\star} (T)}{10}\bigg)^{-4/3} \bigg(\frac{T_{\rm dom}}{T}\bigg)^4 \text{Min}\left[1, \bigg(\frac{T_\text{alp dec}}{T_{\text{mat dom}}}\bigg)^{4/3}\right] \, ,
\\
\nonumber
f_{\rm peak}(T) \approx 1.15 \times 10^{-7} \text{Hz} \times \bigg(\frac{g_\star (T)}{10}\bigg)^{1/2} \bigg(\frac{g_{s\star} (T)}{10}\bigg)^{-1/3} \bigg(\frac{T}{\text{GeV}}\bigg)\text{Min}\left[1, \bigg(\frac{T_\text{alp dec}}{T_{\text{mat dom}}}\bigg)^{1/3}\right]  \, , 
\eea 
Since the largest emission occurs at the annihilation temperature,
one sets $T= T_{\rm ann}$\cite{Hiramatsu:2013qaa}.
After the collapse of a the domain walls, the energy density of the DW is released mostly in the form of axions with a energy density $\rho_a \sim H \sigma$. The axions later decay, but can induce a period of matter domination and dilutes the GW signal, which we accounted for in the last factor. 

Moreover, in the matter domination regime, one needs to take into account the dilution of any other quantity (here the lepton number) from the creation of entropy due to the decay of the axion. 
Following \cite{Scherrer:1984fd, Cirelli:2018iax, Nemevsek:2022anh}, the diluted leptonic yield is 
\begin{equation}
\label{eq:dilution}
    Y_{\Delta L} = Y^0_{\Delta L}\times D\,,\quad \text{with} \quad D = \text{Min}\left[1, 0.57 \frac{g_*(T_\text{ann})}{g_*(T_\text{alp dec})^{1/4}}\frac{\sqrt{M_\text{Pl}\Gamma}\ T_\text{ann}^3}{\Delta V}\right]\,,
\end{equation}
where $Y^0_{\Delta L}$ denotes the leptonic yield as computed in previous sections, without considering an intermediate matter dominated regime.

\section{Baryogenesis via Domain walls: DW-genesis}

 The necessary collapse of the DW network implies that any axionic DW network with a coupling to the lepton or the baryon current would induce lepto/baryogenesis, if there is a lepto/baryon violating operator 
which is effective at the time of the collapse. This claim can be understood as follows: the movement of the DW in the plasma offers the right environment for \emph{spontaneous baryogenesis}. Indeed, due to the coupling between the axion and the lepton current, CP and CPT symmetry are dynamically broken inside the DW. As this is well known, the breakdown of CPT invariance allows to produce baryon or lepton number even in thermal equilibrium. On the top of this, L or B number violation must still be included in the model.  We consider the following right handed neutrino sector 
\bea \label{eq:inte} \mathcal{L} = y_N (  \tilde H \bar L)N_R +  \frac{1}{2} M_N \bar N c_R N_R + \text{h.c.}+ \frac{c_L \partial_{\mu} a}{f_a}  j^\mu_L    \, , \qquad j^\mu_L \equiv \bar L \gamma^{\mu}  L 
\eea
where 
$L$ could designate any lepton flavor, $\tilde H \equiv i\sigma^2 H^\star$, responsible for the neutrino masses via the standard type-I seesaw mechanism~\cite{Minkowski:1977sc,Yanagida:1979as,Mohapatra:1979ia,Schechter:1980gr}. When the DW is sweeping through the plasma, the interaction of Eq.\eqref{eq:inte} in the plasma frame inside the DW takes the form 
\bea
\label{eq:CPT_viol_1}
\qquad \mathcal{L}_{a-L} = \dot \theta j_L^0  \simeq \mu j_L^0 \, , \qquad \qquad 
\dot \theta  \equiv \frac{\dot a_{\rm DW}(t, z)}{f_a} = 
\frac{2 m_a \gamma_w v_w}{\cosh\left[m_a \gamma_w v_w\left(t-t_\text{passage}\right)\right]} \, ,
\eea 
where we used that $j_\mu = j_0(1, 0, 0, 0)$, $t_\text{passage}$ is the time when the region of the plasma under consideration is in the center of the DW. We also introduced $\dot \theta \equiv \dot a/ f_a$ and we will take $t_{\rm passage} \sim t_{\rm ann}$.

\begin{figure}[h!]
    \centering
    \includegraphics[width=.3\linewidth]{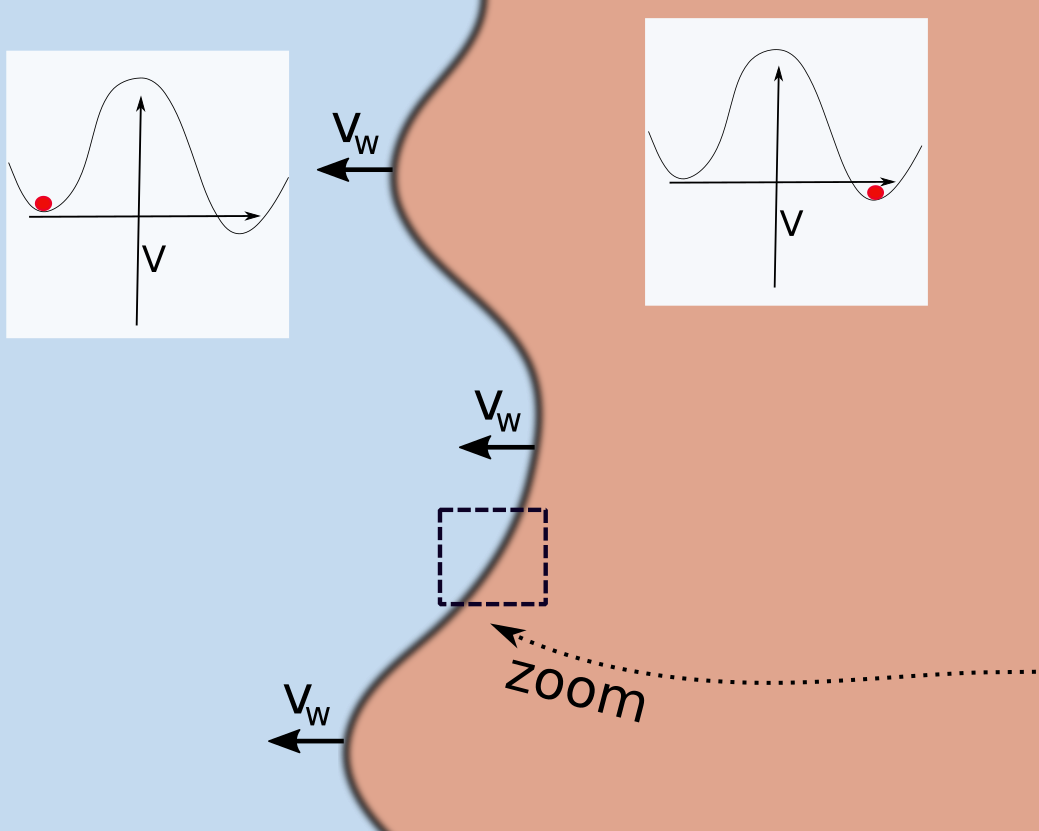}\includegraphics[width=.55\linewidth]{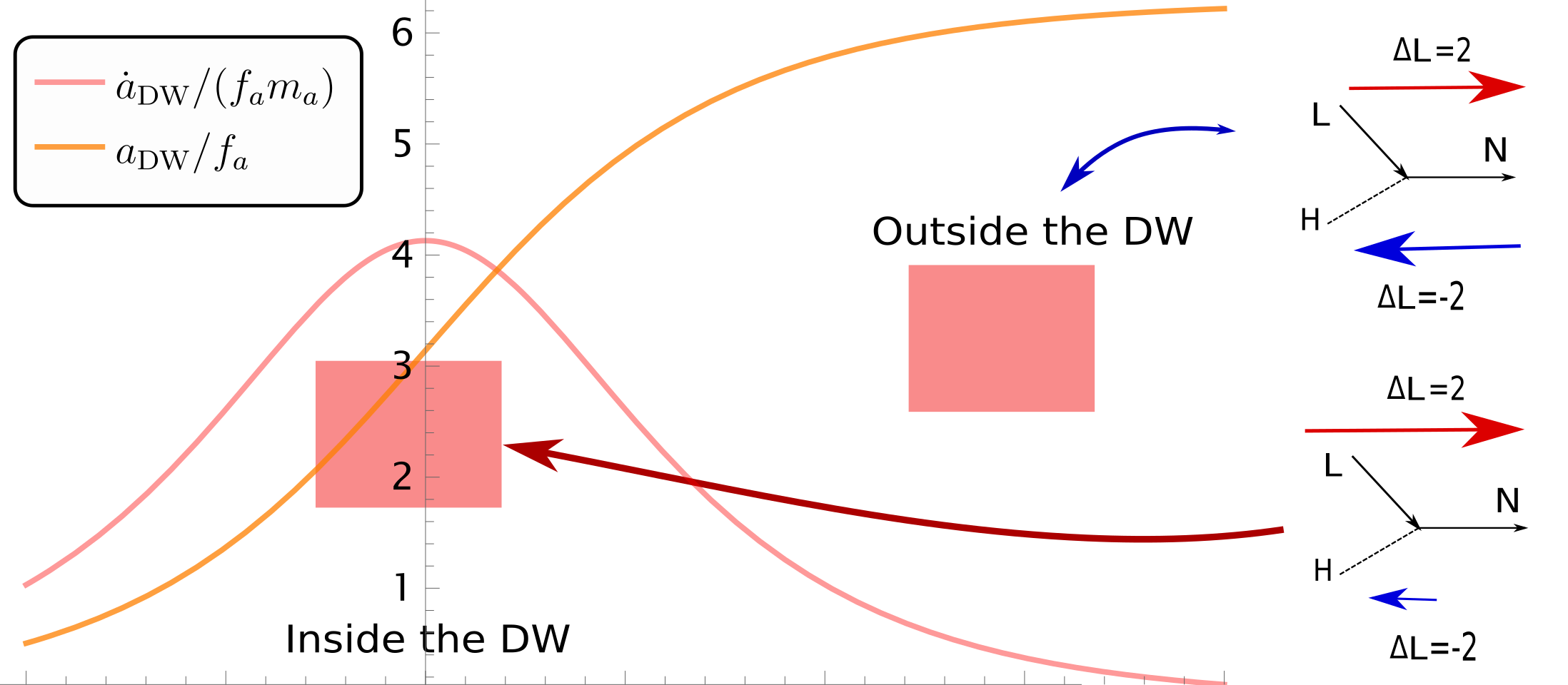}
    
    \caption{\textbf{Scheme of the DW-genesis}: Mechanism discussed in this paper \textbf{Left panel:} Bias-induced collapse of the DW network where the DW moves in a preferred direction. \textbf{Middle panel}: Illustration of the kink DW profile and the induced chemical potential. \textbf{Right panel}: Sketch of the $L$-violating processes active when the DW collapses. Inside the DW, the interactions are biased and the $L$-number is created inside the wall.}
    \label{fig: illustration}
\end{figure}

 As we can see from the first coupling in Eq.\eqref{eq:CPT_viol_1}, the DW acts as a chemical potential $\mu$ for the lepton number. Physically, what happens is that the coupling between the lepton current and the axion, in presence of a varying background field for the axion ($\dot a \neq 0$), can bias the lepton violating interactions and induce and effective chemical potential.
 This process is illustrated in Fig.\ref{fig: illustration}. If the DWs were thick enough, the plasma in the DW would have the time to reach chemical equilibrium and the lepton number would be typically very large. In practice, chemical equilibrium is never reached and one needs to solve Boltzmann equations to quantify the final lepton number. The Boltzmann equations for the lepton number  is
\bea 
\label{eq:lepto_equation}
\frac{dY_{\Delta L}}{dt} = -\bigg( 
\frac{\gamma_D}{n_L^{\rm eq}} (Y_{\Delta L} + Y^{\rm eq}_{\Delta L}(t)) + 2 \frac{\gamma_{2 \to 2} }{n_L^{\rm eq}} \bigg( Y_{\Delta L} + Y^{\rm eq}_{\Delta L}(t)\bigg)\bigg) \, , \quad Y^{\rm eq}_{\Delta L}(t) \equiv \frac{n_{L}^{\rm eq}}{s} \frac{2c_L\dot a}{f_a T}, \quad Y_{\Delta L} \approx 2\frac{n_{L}^{\rm eq}}{s}  \xi_L
\eea 
where we defined the yield via the usual $Y\equiv n/s$, where $s$ is the entropy density
and where $g_{L} = 2$ are the number of degrees of freedom of the lepton.
The rates in Eq.\eqref{eq:lepto_equation} can be obtained from \cite{Buchmuller:2002rq}(see also\cite{Davidson:2008bu, Buchm_ller_2005, Giudice:2003jh})
\begin{align}
\frac{\gamma_D}{n_L^{\rm eq}} =  \frac{M_N^2}{ T^2} K_1(z)y_N^2\frac{M_N}{16 \pi}  \,  \quad 
\frac{\gamma^{\Delta L = 2}_{2 \to 2} }{n_L^{\rm eq}} \approx \Gamma^{\Delta L = 2}_{2 \to 2} \approx \Gamma_{HL \to H^cL^c}+ \Gamma_{LL \to H^cH^c}\,. 
\end{align}

This computation is only valid if the plasma can thermalise in the DW, which amounts to require $\Gamma_{2 \to 2} \gg 1/L_{\rm DW}$. 

The asymmetry can however be washed out if different regions of the plasma produce opposite lepton number. Such a cancellation can occur in the DW-genesis scenario. Recalling that the sign of the asymmetry is determined by the change of $a/f_a$ across the DW. However, if we consider a specific $U(1) $ symmetry and $N_{\rm DW} = 2$ for simplicity, there exists in general two possible different DWs: i) $0 \to \pi$ and ii) $\pi \to 2\pi$. Assuming that the minimum in $\pi$ is favored by the bias, the collapse of the $0 \to \pi$ DW will induce a positive asymmetry while the collapse of $\pi \to 2\pi$ will induce a negative asymmetry, which will eventually cancel each other after diffusion of the baryon number. However, even in this case, 
we do not expect an exact cancellation of the asymmetry, for the following reasons: due to the bias, the tension of the two DWs and their velocity in the plasma will slightly differ, by a quantity proportional to the $\Delta V/V_0$. Consequently, the asymmetry (with an opposite sign) produced by each population of DWs will differ by a quantity proportional to $\Delta V/V_0$, the suppression is given by $\mathcal{O}(1-10) \frac{\Delta V}{V_0}$. In such situation, the \emph{maximal} asymmetry that can be realised (by taking $T_{\rm dec} \sim T_{\rm ann} \sim M_N/10$) is 
\bea
\label{eq:analytics_max_suppressed}
Y^{\text{cancellation}}_{\Delta L}\big|_{\text{max}} 
\approx
10^{-10}
\times 
\mathcal{O}(10)
\left(\frac{T_{\rm ann}}{10^{13} \, \text{GeV}} \right)
\left(\frac{\Delta V/V_0}{10^{-4}}\right) D
\\ 
Y^{\text{no cancellation}}_{\Delta L}\big|_{\text{max}} 
\approx
10^{-10}
\times 
\mathcal{O}(10)
\left(\frac{T_{\rm ann}}{10^{9} \, \text{GeV}} \right)
 D
\eea
where 
$D$ is the dilution factor in Eq.\eqref{eq:dilution}, 
and we normalized $T_{\rm ann}$ around the highest possible value and we considered the two possible case where the cancellation is present or not.

\section{Results and discussion}

\begin{figure}[h!]
    \centering
    \includegraphics[width=.52\linewidth]{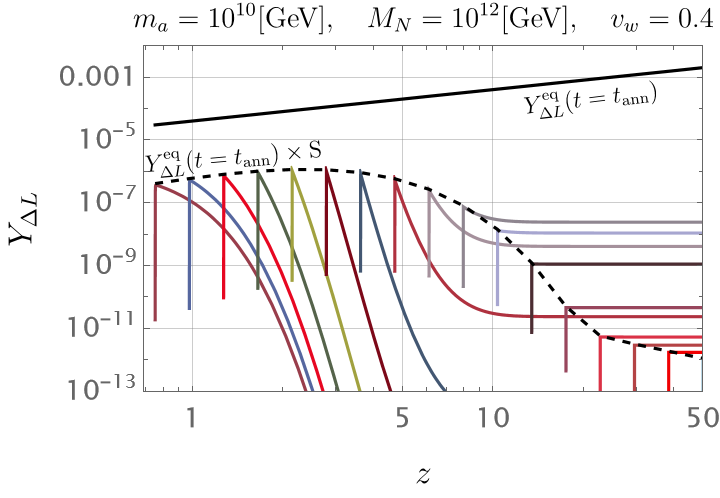}
     \includegraphics[width=.4\linewidth]{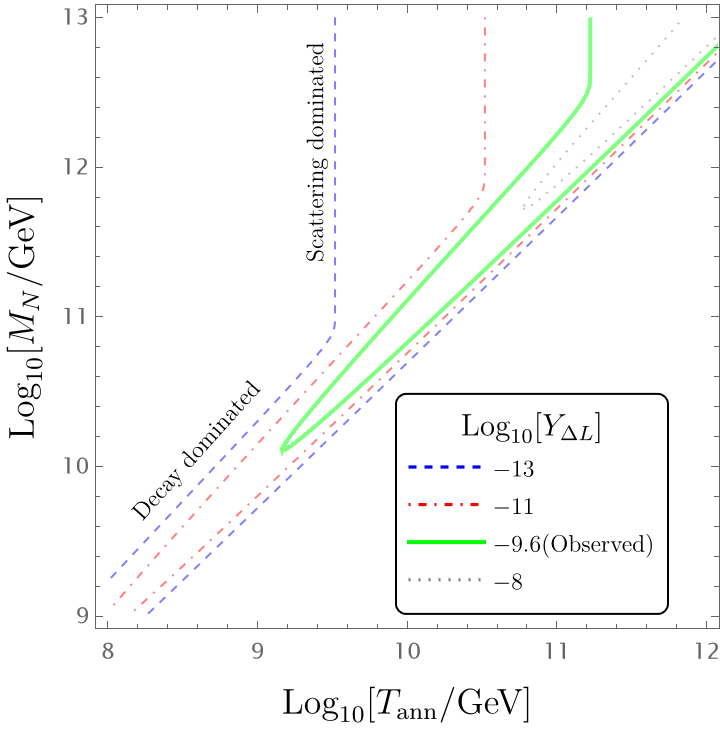}
    \caption{\textbf{Left panel}: trajectories in $z \equiv M_N/T$, the initial growth is due to the passage of the DW and the following decay is due to the wash-outs. \textbf{Right Panel}: Contours of the final abundance $Y_{\Delta L}^{\rm final}$,  the lepton asymmetry in the parameter space $T_{\rm ann} \text{ vs } M_N$. The Green contour illustrates the region matching the observed asymmetry today.
    }
    \label{fig: T_ann}
\end{figure}

\begin{figure}[t!]
\centering
\includegraphics[width=.46\linewidth]{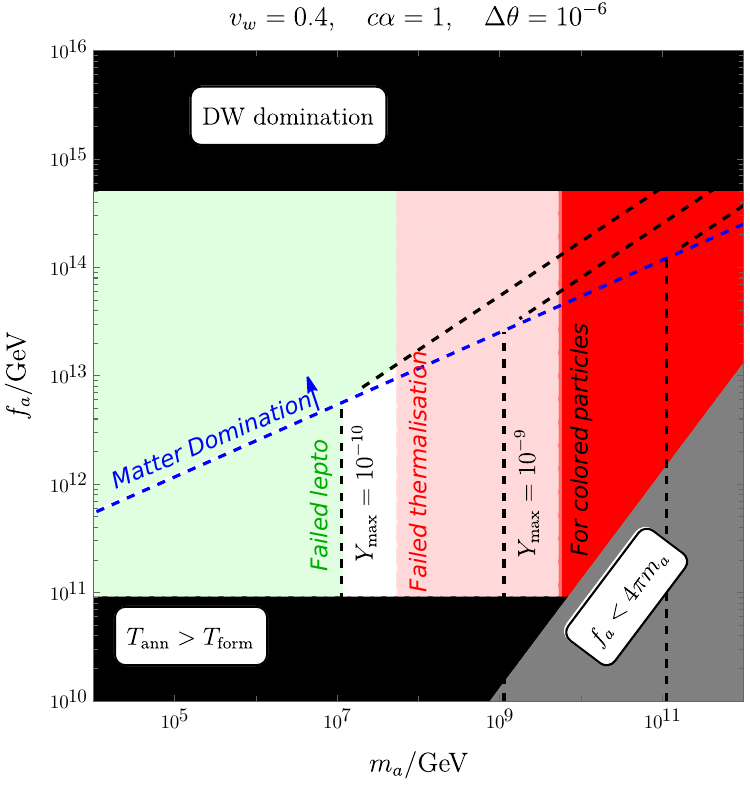}
\includegraphics[width=.48\linewidth]{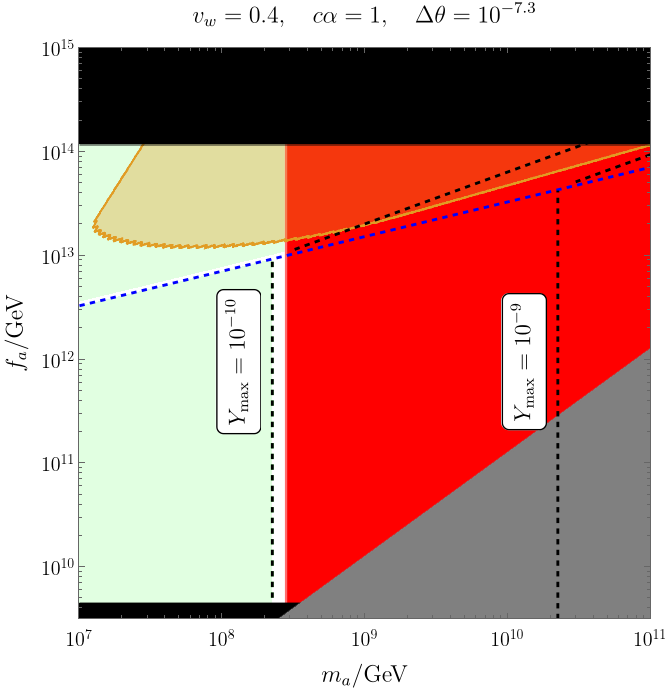}
    \caption{
    \textbf{Left panel}: Plot of the parameter space available for DW leptogenesis for $\Delta \theta = 10^{-6}$.  The Green region does not allow possible leptogenesis because the yield is always too small to match the observed baryon number, and the Red regions exclude the parameter space where the leptonic plasma cannot thermalise in the DW. The dashed Blue curves show the region where the axions emitted by the DWs enter in matter domination, inducing a dilution of the baryon yield, where we set $c \alpha \simeq 1$ in the ALP decay width for concreteness. The viable region for leptogenesis is displayed in White. We emphasize that no region of this plot is observable via a GW signal from the Einstein Telescope.  The maximal value of the asymmetry can be read from the dashed Black contours. \textbf{Right panel}: Same plot for $\Delta \theta = 10^{-7.3}$. We see that even in the optimal case, successfull baryogenesis is in tension with observable GW signal.
    }
    \label{fig:ALPpara}
\end{figure}

On Fig.\ref{fig: T_ann}, we show the solutions of the Eq.\eqref{eq:lepto_equation} \emph{assuming no cancellation due to opposite DWs}. On the Left panel, we show the trajectory of the lepton number as a function of temperature. We observe a first very steep growth due to the passage of the DW and then a slower decay of the asymmetry due to the wash-out. On the Right panel, we show the region fulfilling the observed baryon abundance. This is consistent with the inspection of Eq.\eqref{eq:analytics_max_suppressed}, which shows that in the absence of cancellation, DW-genesis can be efficient for temperature as low as $T \sim 10^{9}$ GeV. Including the cancellation would increase this minimal temperature. On the left panel of Fig.\ref{fig:ALPpara}, we present the parameter space (in the case in which there is no cancellation) where DW-genesis can explain the observed baryon abundance.

We finally find that  
the GW signal emitted by the DW is strongly diluted by the matter domination 
due to axions 
and is eventually not observable at the Einstein telescope\cite{Moore:2014lga,Sathyaprakash:2012jk, Maggiore:2019uih}
in the realisation discussed in this paper. This can be observed on the right panel of Fig.\ref{fig:ALPpara}, where we observe that successful baryogenesis and observable gravitational wave are in tension. This however motivates further model building to investigate how generic is this conclusion.

\section*{Acknowledgments}

MV is supported by the ``Excellence of Science - EOS'' - be.h project n.30820817, and by the Strategic Research Program High-Energy Physics of the Vrije Universiteit Brussel.

\bibliographystyle{JHEP}
{\footnotesize
\bibliography{biblio}}

\end{document}